\documentclass[showpacs,aps,prl,twocolumn,floatfix,superscriptaddress]{revtex4}
\usepackage{graphicx}
\usepackage{epsfig}
\usepackage{wasysym}
\usepackage{CJKutf8}

\begin{document}
\begin{CJK*}{GBK}{}
\title{Tunable anisotropic superfluidity in an optical kagome superlattice}
\author{Xue-Feng Zhang}   
\affiliation{Physics Department and Research Center OPTIMAS, Technical University of Kaiserslautern, 67663 Kaiserslautern, Germany}
\author{Tao Wang}
\affiliation{Physics Department and Research Center OPTIMAS, Technical University of Kaiserslautern, 67663 Kaiserslautern, Germany}
\affiliation{Department of Physics, Harbin Institute of
Technology, Harbin 150001, China} 
\author{Sebastian Eggert}
\affiliation{Physics Department and Research Center OPTIMAS, Technical University of Kaiserslautern, 67663 Kaiserslautern, Germany}
\author{Axel Pelster}
\email{axel.pelster@physik.uni-kl.de} 
\affiliation{Physics Department and Research Center OPTIMAS, Technical University of Kaiserslautern, 67663 Kaiserslautern, Germany}

\begin{abstract}
We study the phase diagram of the Bose-Hubbard model on the kagome lattice
with a broken sublattice symmetry.  Such a superlattice structure can naturally be created and
tuned  
by changing the potential offset of one sublattice in the 
optical generation of the frustrated lattice.
The superstructure
gives rise to a rich quantum phase diagram, which 
is analyzed by 
combining Quantum Monte Carlo simulations with the
Generalized Effective Potential Landau Theory. 
Mott phases with non-integer filling and
a characteristic order along stripes are found, which show a transition to a superfluid 
phase with an {\it anisotropic} superfluid density.  
Surprisingly, the direction of the superfluid anisotropy 
is changing 
between different symmetry directions as a function of the particle number or
the hopping strength. Finally, we discuss characteristic signatures of
anisotropic phases in 
time-of-flight
absorption  measurements.
\end{abstract}


\maketitle
\end{CJK*}

Ultracold atoms in optical lattices are prominently used to simulate
many-body systems in condensed matter
physics \cite{qop1,qop2,qop3,qop4,qop5}. One of the most striking 
experiments is the
Mott insulator--superfluid quantum phase transition of ultracold
bosons in optical lattice built with counter-propagating
lasers \cite{bh1}. It can be described by the seminal Bose-Hubbard
model \cite{bh0,bh2}, where each parameter is precisely adjustable
in the experiment.   With the rapid advances in experimental 
techniques \cite{toolbox} the many-body physics can now be analyzed 
on more complex lattice geometries.
On the one hand, the lattice symmetry can be reduced by adding 
additional lasers or tuning
their relative strength, leading to a superlattice structure \cite{sp1,sp2,sp3,sp4}, 
which can give rise to insulator phases with fractional fillings \cite{sp5,sp6,sp7,sp8,tao}.
On the other hand, it is also possible to {\it enhance} the residual entropy of the 
many-body system by using frustrated lattices, which have recently been realized using 
sophisticated optical techniques \cite{windpassinger,sengstock,klattice}.

Theoretically, many interesting phases have been predicted in frustrated lattices
such as spin-liquids \cite{wen,sl1,sl2,sl3,sl4,sl5,sl6}, valence bond solids
\cite{kl1}, string excitations \cite{kl0}, ordered metals \cite{tocchio}, 
chiral fractional edge states \cite{kl2}, and supersolids \cite{tri_sc1,tri_sc2,tri_sc3,sellmann}.
Unfortunately, however, in all these scenarios longer range interactions beyond 
the on-site Bose-Hubbard model 
are assumed, which require dipolar interactions and are experimentally 
much harder to handle.
On the other hand, the intriguing interplay between a 
superlattice and the kagome lattice has never been explored
before for the on-site Bose-Hubbard model.  
This is surprising since a superlattice structure can be created naturally
in optically generated kagome lattices, and insulating phases with fractional filling occur
without the need of longer range interaction as discussed below.
We now analyze the detailed phase diagram of 
the Bose-Hubbard model on the kagome lattice with a tunable superlattice structure.
In addition to the fractionally filled ordered phases in the quantum phase diagram, 
the most striking features are found 
in the unusual properties of the
Bose condensed phase, where the anisotropic superfluid density spontaneously picks 
a preferred direction depending on filling and interaction.
This leads to a characteristic signature in time-of-flight experiments.

\begin{figure}[t]
\includegraphics[width=0.5\textwidth]{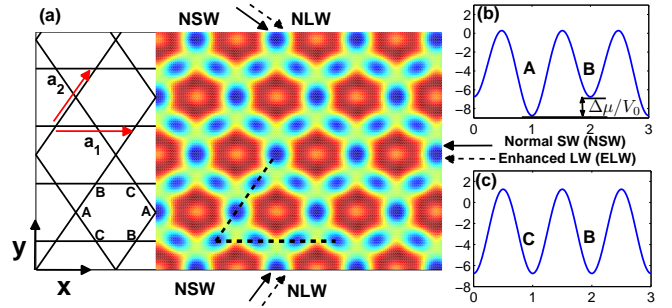}
\caption{(a) Potential from Eq.~(\ref{V}) of the optical kagome superlattice
using an enhanced LW laser in the $x$-direction with 
$\gamma=1.5$. 
Potential along cuts (dashed lines) in the b) $a_2$-direction and c) $a_1$-direction, respectively, 
showing the resulting potential offset $\Delta\mu$ for sublattice A. 
\label{model}}
\end{figure}

Let us first consider the optical generation of a kagome lattice, which recently has 
been achieved
in experiment by using standing waves from a long wavelength 1064~nm (LW) laser 
and from a short wavelength 532 nm (SW) laser, which are counterpropagating from three 
$120^\circ$ directions \cite{klattice}.  
The superposition of the corresponding two triangular lattices results
in a kagome lattice 
if the laser strengths 
are exactly equal from all directions.  Any slight variation of this
setup results in a superlattice structure, which of course can in turn be
used as an additional tunable parameter.  For example, enhancing the 
potential from the LW laser in the $x$-direction by a factor $\gamma=V_E/V_0>1$
results in a combined optical potential 
\begin{eqnarray}
V_c/V_0&=&\gamma^2-1
+4\gamma\cos(\sqrt{3}kx)\cos(ky)+2\cos(2ky)\nonumber \\
&&-2\cos(4ky)-4\cos(2\sqrt{3}kx)\cos(2ky)\,,
\label{V}
\end{eqnarray}
where $k=\sqrt{3}\pi/2 \lambda_{\rm LW}$ in units of the 
longer wavelength $\lambda_{\rm LW} = 1064\rm nm$.  As depicted in Fig.~\ref{model}
this potential leads to an offset of $\Delta\mu=4(\gamma-1)V_0>0$ on one
of the sublattices A.  Note, however, that this offset preserves the parity
symmetry along the $x$- and the $y$-direction and does not increase the
unit cell of the kagome lattice, which contains three sites.

Interacting bosons on this lattice can be represented by Wannier states, 
which leads to the well-known Bose-Hubbard model for the description 
of the lowest band in second quantized language
\begin{eqnarray}
\nonumber H&=&-t\sum_{\langle
i,j\rangle}(\hat{a}_{i}^{\dag}\hat{a}_{j}+\hat{a}_{i}\hat{a}_{j}^{\dag})+\frac U 2 \sum_{i}\hat{n}_{i}(\hat{n}_{i}-1)\\
&&-\mu \sum_{i}\hat{n}_{i}-\Delta \mu \sum_{i\in A}\hat{n}_i\,.
\label{EBH}
\end{eqnarray}%
where the nearest-neighbor hopping amplitude $t$ and the onsite interaction $U$
are tunable parameters, which depend on the scattering cross-section and the potential depth 
$V_0$ \cite{qop4}.
In principle, the potential shift $\Delta\mu$ also affects the Wannier states and hence other 
parameters in Eq.~(\ref{EBH}), but for 
reasonably small values of $\Delta\mu$
these higher order corrections can be neglected since 
they preserve the symmetry of the problem.
The chemical potential $\mu$  is used to 
tune the particle number in the grand-canonical ensemble.
In the following we will use 
the stochastic cluster series 
expansion  algorithm \cite{sse1,sse2,scse} for unbiased quantum Monte
Carlo (QMC) simulations of this model.  
In addition the Generalized Effective Potential
Landau Theory (GEPLT) provides an analytic method to estimate
the phase boundaries in an expansion of the hopping parameter $t/U$ \cite{tao,santos}. 
\begin{figure}[t]
\includegraphics[width=0.5\textwidth]{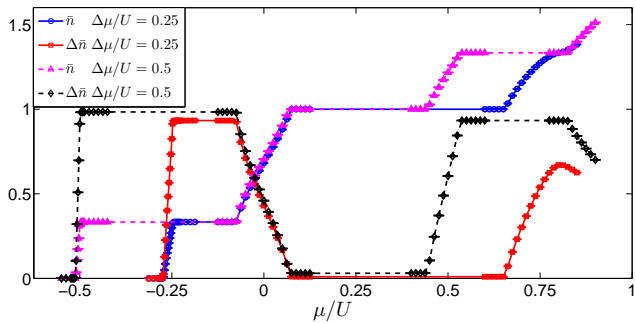}
\caption{Average density $\overline{n} =(n_{\rm A} + n_{\rm B} + n_{\rm C})/3$ and density 
difference $\Delta n = n_{\rm A} - (n_{\rm B}  + n_{\rm C})/2$
versus $\mu$ from QMC with 
$T=U/300$ and $N=243$ sites at $t/U=0.025$ for different offsets $\Delta\mu/U$.
 \label{rho}}
\end{figure}

\begin{figure}[t]
\includegraphics[width=0.5\textwidth]{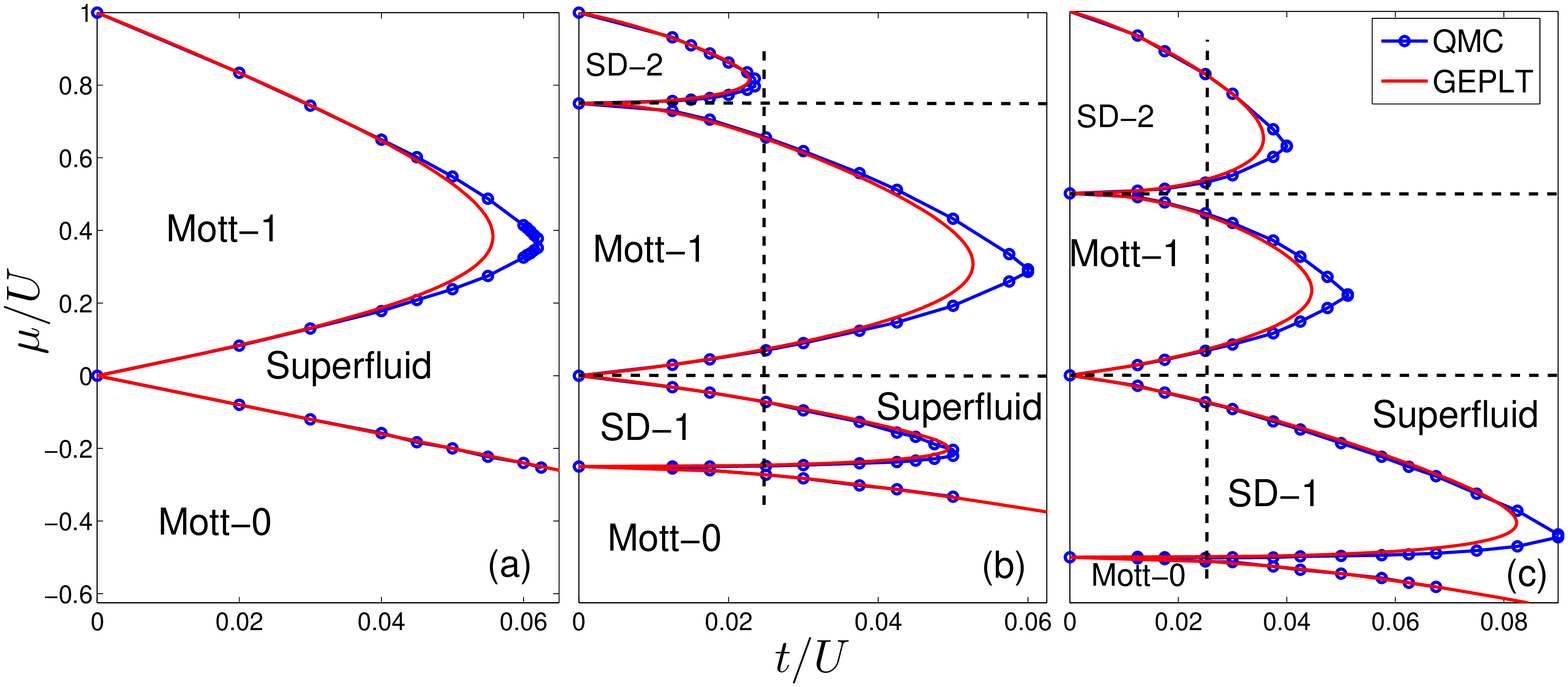}
\caption{Quantum phase diagram extrapolated to the thermodynamic limit obtained
from QMC (blue) and multi-component GEPLT in second order in $t/U$ (red) 
at (a) $\Delta\mu/U=0$, (b) $\Delta\mu/U=0.25$, and (c)
$\Delta\mu/U=0.5$.  The vertical lines indicate the parameter ranges in Figs.~\ref{rho} and \ref{rhos}
while the horizontal lines are used in Fig.~\ref{rhocomp}
\label{phase}}
\end{figure}

For vanishing hopping amplitude $t$ in the atomic limit, the competition between $U$
and $\Delta\mu$ can induce several incompressible insulating
phases. When $\mu$ is less
than $-\Delta\mu$, no site is occupied, and the Mott-$0$ phase
is the energetically favored state. For a larger chemical potential
 $-\Delta\mu< \mu <0$, 
only sublattice A will be occupied with one boson per site while the other sites remain empty.
This phase is therefore $1/3$ filled with an order in the form of occupied 
horizontal stripes.  Such a 
$1/3$ striped density phase (SD) can also occur spontaneously in the 
extended Hubbard model when nearest and next-nearest interactions are included \cite{kl2}.
However, longer range interactions are notoriously difficult in optical lattices, so 
that the proposed superlattice is a convenient tool to study this phase.
For positive values of 
$\mu > 0$, the system enters the familiar 
uniform Mott-1 insulator with filling factor one. Continuing this analysis for larger 
$\mu$, we deduce that SD-$n$ phases with fractional filling factor
$n-2/3$ occur for
$U(n-1)-\Delta\mu < \mu < U(n-1)$, which are separated by Mott-$n$ insulators with
integer filling $n$ for
$U(n-1) < \mu < Un-\Delta\mu$.

Both the integer filled Mott-$n$ phases and the fractional SD-$n$ phases remain stable
for small finite hopping $t$.  
As shown in Fig.~\ref{rho} for $t=0.025~U$ there are 
plateaus of the average density $\overline{n} =(n_{\rm A} + n_{\rm B} + n_{\rm C})/3$
as a function of chemical potential, which 
are characteristic of those incompressible phases.
In the fractionally filled SD-$n$ phases
the density difference $\Delta n = n_{\rm A}-(n_{\rm B}+n_{\rm C})/2$
between the sublattices shows plateaus with a value that is slightly reduced from
unity due to virtual quantum excitations.
The plateau states are separated by compressible phases, which are characterized
by a finite superfluid density, i.e.~an off-diagonal order with a spontaneously broken
U(1) gauge symmetry which will be analyzed in more detail below.

The corresponding phase diagram is mapped out in Fig.~\ref{phase} using 
large scale QMC simulations.   The second order 
GEPLT approximation is much less demanding and agrees 
quite well with the QMC data, except near the tips of the Mott lobes. 
With increasing offset $\Delta \mu$ the
fractionally filled SD phases extend over a larger range not only in the chemical 
potential $\mu$ but also in the 
hopping $t$.  In fact, the SD-1 phase for $\Delta \mu = 0.5~U$
is remarkably stable up to larger values of  hopping $t$ than the 
uniform Mott-1 phase. 
The transitions to the superfluid phase are always of second order and can 
be understood in terms of additional condensed particles (holes) on top of the
Mott states as the chemical potential is increased (decreased).

One interesting detail in the phase diagram in Fig.~\ref{phase} is the drastic dependence 
on $\Delta \mu$ of the 
shape of the Mott-0 phase transition line in the limit of small hopping, 
which changes from linear behavior 
$\mu(t) = -4t$ for $\Delta \mu = 0$ to quadratic 
behavior $\mu(t) = -\Delta\mu-8t^2/\Delta\mu$ for large $\Delta \mu$.
The linear dependence for $\Delta \mu=0$
can be understood from a competition of chemical potential with 
the kinetic energy, analogously to the quantum melting on the triangular 
lattice \cite{tri_sc1}.  For finite $\Delta\mu$, on the other hand, the melting of the
 Mott-0 phase takes place by additional particles 
on the sublattice A only, which is not connected by any first-order hopping processes.
In the limit of small $t$, the kinetic energy of those particles is therefore 
determined by the second order 
hopping coefficient 
$\tilde t = t^2/\Delta\mu$, which explains the quadratic behavior of the phase boundary.
The exact shape of the Mott-0 transition 
$\mu=-\Delta\mu/2-t-\sqrt{\Delta\mu^2-4t\Delta\mu+36t^2}/2$ can be determined 
from the 
single particle energy on the superlattice.

\begin{figure}[t]
\includegraphics[width=1.\columnwidth]{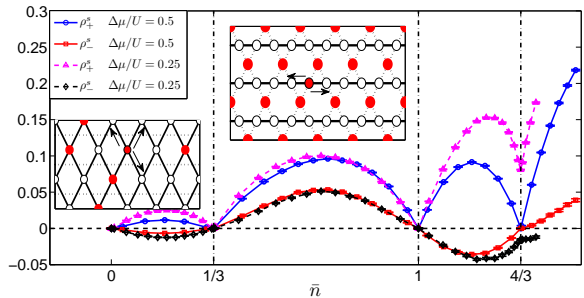}
\caption{Total superfluid density $\rho^{\rm s}_+=(\rho_1^{\rm s}+\rho_2^{\rm s})/2$ and 
superfluid density difference $\rho_-^{\rm s}=\rho_1^{\rm s} - \rho_2^{\rm s}$ versus filling 
$\bar n$ from QMC 
for $\beta U=1200$ and
$N=243$ at $t/U=0.025$. Inset: schematic illustration of the different
mechanisms for positive and negative anisotropy parameters.
\label{rhos}}
\end{figure}

We now turn to the analysis of the order parameter in the superfluid phase. 
In the QMC simulations we determine the superfluid density along the 
lattice vector direction $\vec{a}_1 = (1,0)$ 
using the winding number $\rho_1^{\rm s} = \langle W_1^2\rangle/4\beta t$
and analogous for $\rho_2^{\rm s}$ along the lattice direction $\vec{a}_2 = (1,\sqrt{3})/2$
\cite{winding}.    We use a system with $N=243$ sites and periodic boundary conditions
with $L=9$ unit cells in both the $\vec{a}_1$ and $\vec{a}_2$ directions, which ensures that 
$\rho_1^{\rm s}=\rho_2^{\rm s}$ for the perfect kagome lattice. 
Note, that in general the superfluid density is a 
response {\it tensor} with four elements $\rho_{xx}^{\rm s}, \, \rho_{xy}^{\rm s}, \, \rho_{yx}^{\rm s},\, \rho_{yy}^{\rm s}$ 
in the $x$-$y$-coordinate system \cite{ueda}.  Due to reflection
symmetry the off-diagonal elements $\rho_{xy}^{\rm s}=\rho_{yx}^{\rm s} = 0$ must vanish.
The relation to the superfluid densities along the 
lattice vectors is given by $\rho_1^{\rm s}=\rho_{xx}^{\rm s}$ and $\rho_2^{\rm s} = (\rho_{xx}^{\rm s}+3\rho_{yy}^{\rm s})/4$.

In order to analyze a possible anisotropy we consider the average superfluid density 
$\rho_+^{\rm s} = (\rho_1^{\rm s}+\rho_2^{\rm s})/2$
and the difference $\rho_-^{\rm s} = \rho_1^{\rm s} -\rho_2^{\rm s}$ between the two lattice vector directions
in Fig.~\ref{rhos} as a function of filling $\bar n$.
For finite offsets $\Delta \mu =0.25~U$ and $\Delta \mu =0.5~U$ the superfluid density
is indeed anisotropic, but surprisingly also changes the preferred 
direction with increasing filling $\bar n$.
For low densities just above the Mott-0 phase the superfluid density is 
dominated by virtual hopping processes between the A sublattice.
As illustrated in the left inset of Fig.~\ref{rhos} this virtual hopping process
is not possible along the lattice vector $\vec{a}_1$, which leads to an
anisotropic superfluid density with $\rho_1^{\rm s} < \rho_2^{\rm s}$.

When the filling reaches $\bar n=1/3$ the superfluid density drops to zero in the SD-1 phase 
as expected, but then 
shows the opposite anisotropy $\rho_1^{\rm s} > \rho_2^{\rm s}$ for $\bar n>1/3$, which signals a different
mechanism:  At $\bar n=1/3$ the A sublattice is completely filled, so that for slightly
larger densities $\bar n>1/3$
excess particles on the B and C sublattices are now responsible for
the superfluid density.  As shown in the right inset of 
Fig.~\ref{rhos}, the B and C sublattices correspond to connected 
chains along
the $\vec{a}_1$ direction, which are disconnected by occupied A sites. This immediately 
explains why $\rho_1^{\rm s} > \rho_2^{\rm s}$ in this case. 

According to this analysis, 
positive anisotropies $\rho_-^{\rm s} > 0$ are therefore a hallmark of an
off-diagonal U(1) order parameter coexisting with a striped density order 
of a filled sublattice A.
This situation is reminiscent of a supersolid where a stable density order exists
on one filled sublattice and excess particles contribute to the 
superfluidity \cite{tri_sc1}, with the main difference that 
in supersolids both the U(1) symmetry and the translational symmetry are
{\it spontaneously} broken.  Normally supersolid phases require
longer range interactions beyond on-site, which are experimentally difficult to achieve.
The creation of 
supersolid-like regions by introducing a superlattice is experimentally straight-forward, 
however.  Similar to the ordinary supersolid, the supersolid-like regions considered
here are also only stable for relatively small hopping, while for larger hopping
the "ordinary" superfluid behavior dominates as we will see below.

As long as the hopping $t$ is sufficiently small the
alternation of anisotropies between Mott and SD phases continues as the density is increased
due to the same reasoning as above.
However, this is not the full story since for larger hopping $t$ or larger filling $\bar n$ the
Mott and SD phases are not stable, so it is not clear where the different regions of
positive and negative $\rho_-^{\rm s}$ are separated.  Indeed as shown in Fig.~\ref{rhos},
the superfluid density does not drop to zero for $\bar n=4/3$ and $\Delta\mu/U=0.25$, since
the corresponding line is just outside the lobe of the SD-2 phase as 
shown in Fig.~\ref{phase}(b).
Also the anisotropy no longer changes sign.  
We find that in the limit of large hopping $t$ the overall density becomes irrelevant. The 
sublattice A remains slightly more occupied for all values of $\bar n$ and $t$. Since particles
on the A sublattice hardly hop in the $\vec{a}_1$ direction, this leads
to $\rho_1^{\rm s}<\rho_2^{\rm s}$ in the weak coupling limit $t>U$.  We call this behavior the "ordinary"
superfluid, in contrast to the supersolid-like regions of 
the positive anisotropy $\rho_-^{\rm s} >0$, which are basically confined between the lobes of the
SD-n and Mott-n phases. 

\begin{figure}[t]
\includegraphics[width=0.5\textwidth]{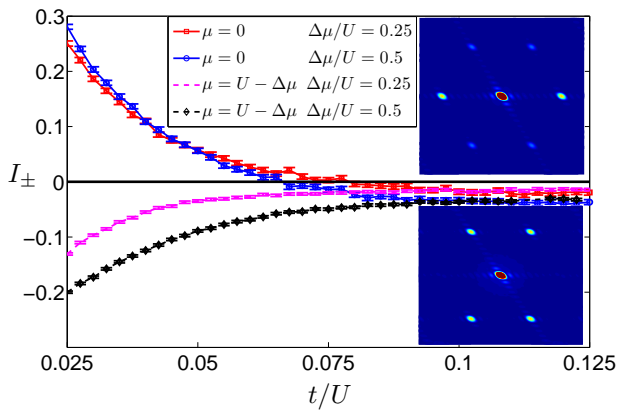}
\caption{Superfluid anisotropy parameter $I_{\pm}=(\rho_1^{\rm s}-\rho_2^{\rm s})/(\rho_1^{\rm s}+\rho_2^{\rm s})$ 
as a function of $t/U$ from QMC for $\beta U=2000$ and $N=432$ in the parameter range indicated by the horizontal lines in Fig.~\ref{phase}. Insets: 
QMC simulations of the TOF image for  $t/U=0.0375$, $\beta U=800$, $N=243$, 
$\Delta\mu/U=0.5$, and $\mu/U=-0.175$ (top) and  $\mu/U=0.425$ (bottom). 
\label{rhocomp}}
\end{figure}

To analyze the crossover between different anisotropy regions we show the
normalized anisotropy parameter
$I_{\pm}=(\rho_1^{\rm s} -\rho_2^{\rm s})/(\rho_1^{\rm s}+\rho_2)^{\rm s}$ as a function
of $t/U$ for different values $\mu$ and $\Delta\mu$ in Fig.~\ref{rhocomp}. 
For small hopping the anisotropy parameter
$I_{\pm}$ is positive in supersolid-like regions ($\mu=0$) and negative
between the Mott-$1$ and SD-$2$
phase ($\mu=U-\Delta\mu$) as discussed above.  For larger $t/U$ the anisotropy parameters
approach 
small negative values in all cases, corresponding to the ordinary superfluid.
According to the analysis above, the sign-change of $I_\pm$ as a function of $t$
coincides with the delocalization of the particles on the A sublattice, which start to 
contribute to the superfluid density in the $\vec{a}_2$ direction.  This behavior 
can be interpreted as a continuous "melting" of the supersolid-like phase to the ordinary superfluid, 
reminiscent of 
the melting of the sublattice order in an interaction driven supersolid \cite{tri_sc1}.

Anisotropic superfluid densities appear in a variety of different systems such as
dipolar Bose-Einstein condensates with disorder \cite{krumnow,nikolic,ghabour}, 
spin-orbit coupled Fermi gases \cite{devreese}, coupled spin dimer systems \cite{strassel}, 
and systems with rectangular shape \cite{sf}. 
However, an anisotropic superfluidity which is tunable by the isotropic hopping $t$ and
changes sign when the order on one sublattice melts has not been discussed before 
to our knowledge.

The observation of the superfluid-Mott transition by time-of-flight (TOF) experiments
has been pioneered many years ago \cite{bh1}.  The TOF absorption picture measures
the momentum distribution $S(\textbf{Q})/N=\langle |\mathop{\sum}_{k=1}^N a_k^+
e^{\emph{i} \textbf{Q} \cdot\textbf{r}_k}|^2\rangle/N^2$ and turns out to also show a clear
signature of the anisotropy parameter.  To demonstrate this effect, we used 
a QMC technique for calculating the  off-diagonal long-range
correlation function during the loop update \cite{off}, which allows a direct simulation of
the TOF absorption signal. As shown in Fig.~\ref{rhocomp} 
for $\rho_-^{\rm s} >0$ (upper inset) and for $\rho_-^{\rm s} < 0$ (lower inset) 
the TOF images 
display a clear signature of the anisotropy, which can be used for straight-forward 
measurements of the melting
from supersolid-like to ordinary superfluid states.

In conclusion, we analyzed ultracold bosons in a kagome
superlattice, which can be created and tuned  by enhancing the long wavelength laser in one
direction based on recent progress for creating highly frustrated lattices \cite{klattice}. 
By using numerical QMC simulations and the
Generalized Effective Potential Landau Theory, we obtained the entire 
quantum phase diagram including Mott phases and fractionally filled charge density
phases.  In the superfluid phase an anisotropic superfluid density is found,
which changes direction as the overall density or the hopping is changed.
By tuning the hopping it is possible to induce a continuous {\it melting} from
a supersolid-like state with a filled sublattice A and positive anisotropy
parameter $\rho_-^{\rm s}>0$ to an ordinary superfluid phase, which generically
is characterized by a negative anisotropy parameter $\rho_-^{\rm s}<0$.
Both the fractionally filled insulating phases and supersolid phases have received
much attention by using models with longer-range 
interactions \cite{kl1,kl0,tocchio,kl2,tri_sc1,tri_sc2,tri_sc3}.  Using the 
superlattice structure proposed in this work these phases become experimentally 
much more accessible by a simple laser setup instead of using dipolar interactions.  
Moreover, the characteristic signature 
of those effects can be measured in straight-forward TOF absorption experiments,
without the need of single site resolution.
In particular,  
by implementing off-diagonal measurements in QMC loop updates, it was possible to 
simulate TOF flight images which show a clear signature of the anisotropic 
superfluid density and the change of its direction, when the melting takes place.

\begin{acknowledgments}

X.-F. Zhang thanks for discussions with D. Morath about TOF calculations 
and with Y.C. Wen about 
superlattices. This work was supported by the Allianz f\"ur
Hochleistungsrechnen Rheinland-Pfalz and by the
German Research Foundation (DFG) via the 
Collaborative Research Center SFB/TR49.

\end{acknowledgments}

\bibliographystyle{apsrev}

\end{document}